\documentclass[prl,aps,twocolumn,showpacs,a4paper]{revtex4}
\usepackage{bm,float,amssymb}
\usepackage[dvips]{graphicx}
\usepackage{amsmath}
\graphicspath{{figures/}}

\begin{document}

\title{Direct Measurement of the Flow Field Around Swimming Microorganisms}

\author{Knut Drescher, Raymond E. Goldstein, Nicolas Michel, Marco Polin, and Idan Tuval}
\affiliation{Department of Applied Mathematics and Theoretical
Physics, University of Cambridge, Cambridge CB3 0WA, UK}

\date{\today}

\begin{abstract}
Swimming microorganisms create flows that influence their mutual interactions
and modify the rheology of their suspensions. While extensively studied
theoretically, these flows have not been measured in detail around any
freely-swimming microorganism. We report such measurements for the microphytes
{\it Volvox carteri} and {\it Chlamydomonas reinhardtii}. The minute
($\sim0.3\%$) density excess of {\it V. carteri} over water leads to a strongly
dominant Stokeslet contribution, with the widely-assumed stresslet flow only a
correction to the subleading source dipole term. This implies that suspensions
of {\it V. carteri} have features similar to suspensions of sedimenting
particles. The flow in the region around {\it C. reinhardtii} where significant
hydrodynamic interaction is likely to occur differs qualitatively from a
``puller'' stresslet, and can be described by a simple three-Stokeslet model.
\end{abstract}


\maketitle

Aided by advances in imaging techniques that allow detailed studies of the
rotating flagella of bacteria \cite{TurnerRyuBerg} and the undulating flagella
of spermatozoa \cite{Friedrich} and algae \cite{Chlamy_Science}, there is now a
general consensus on how mechanical motions of microorganism appendages generate
propulsive forces in a viscous fluid \cite{BrennenWinett}. No such consensus
exists yet on the origins of collective behavior \cite{Dombrowski}, transport
\cite{Underhill,Chlamytracers} and rheological properties of suspensions
\cite{rheology_suspensions}, and the interaction of organisms with surfaces
\cite{Berke,TangCrowdy}. As hydrodynamics surely plays a key role in these
effects, a detailed knowledge of the flow field around \textit{freely swimming}
microorganisms is needed, both in the near-field and far away.  Here we present
the first such measurements.

The linearity of the Stokes equations implies that the far-field flow around a
microorganism can be expressed as a superposition of singularity solutions
\cite{pozrikidis}, with the slowest decaying mode dominating sufficiently far
away. Theories of fluid-mediated interactions and collective behavior typically
assume neutrally buoyant swimmers which exert no net force on the fluid.   The
thrust $T$ of their flagella and the viscous drag on their body are displaced a
distance $d$ apart (often comparable to the cell radius $R$), and balance to
give the far-field flow of a force dipole, or stresslet \cite{Batchelor}, which
decays with distance $r$ as $Td/\eta r^2$, where $\eta$ is the fluid's
viscosity. The contribution from a suspension of such stresslets to the fluid
stress tensor is central to some of the most promising approaches to collective
behavior of microorganisms \cite{Ramaswamy}.  

The force-free idealization of swimmers requires precise density-matching
\cite{Berke} not generally realized in nature. To appreciate the striking
effects of gravity, one need only consider the buoyancy-driven plumes of
bioconvection \cite{PedleyKessler}.   Models of this instability express the
contribution of cells to the Navier-Stokes equations as a sum of force monopoles
(Stokeslets), coarse-grained as a body force proportional to the cell
concentration and gravitational force $F_g$ per cell \cite{PedleyKessler}. As
the flow around a Stokeslet decays as $F_g/\eta r$, it is clear, if not
appreciated previously, that there is a distance $\Lambda\sim Td/F_g$ at which
the nearby stresslet contribution crosses over to the distant Stokeslet regime.
This is one of several crossover lengths relevant to swimmers; for ciliates,
unsteady effects become important on scales smaller than the viscous penetration
depth \cite{Brennen}.  For a given organism, the relevance of the length
$\Lambda$ to a particular physical situation depends on the cell concentration
and the observable of interest.  At low concentrations the Stokeslet form
suffices, but the near field is relevant to cell-cell interactions, especially
in concentrated suspensions \cite{interactions} and to tracer dynamics
\cite{Chlamytracers}. The notion of near field requires distinguishing between
distances $r$ satisfying $R\ll r\ll \Lambda$, where a stresslet description may
hold, and $r\sim R$ where the multipole contributions may not be well-ordered
and the flow topology can differ from that of a stresslet.

A synthesis of tracking microscopy and fluid velocimetry is used here to
quantify the competing force singularities and the near-field flow topology for
the unicellular biflagellate green alga {\it Chlamydomonas reinhardtii}
\cite{harris09} ($R\sim\! 5$ $\mu$m) and its larger relative {\it Volvox
carteri} \cite{Kirkbook}, a spherical alga ($R\sim\! 200$ $\mu$m) which swims by
the action of $\sim\! 10^3$ \textit{Chlamydomonas}-like cells on its surface.
For {\it Volvox} our most significant finding is that the flow field is strongly
dominated by its Stokeslet component, despite a density excess of a mere $\sim
0.3\%$, much smaller than that of common unicellular organisms ($\sim 5-10\%$).
Moreover, the high symmetry of {\it Volvox} results in a leading near-field
correction in the form of a source doublet, and a smaller stresslet.  The flow
around {\it Chlamydomonas} is compatible with a simple ``puller" stresslet only
at distances $\gtrsim 7R$, where the fluid velocity is $\lesssim 1\%$ of the
swimming speed; closer to the cell, the flow topology reflects the finite
separation of the flagellar and body forces. 

\begin{figure}[t]
\begin{center}
\includegraphics*[clip=true,width=0.9\columnwidth]{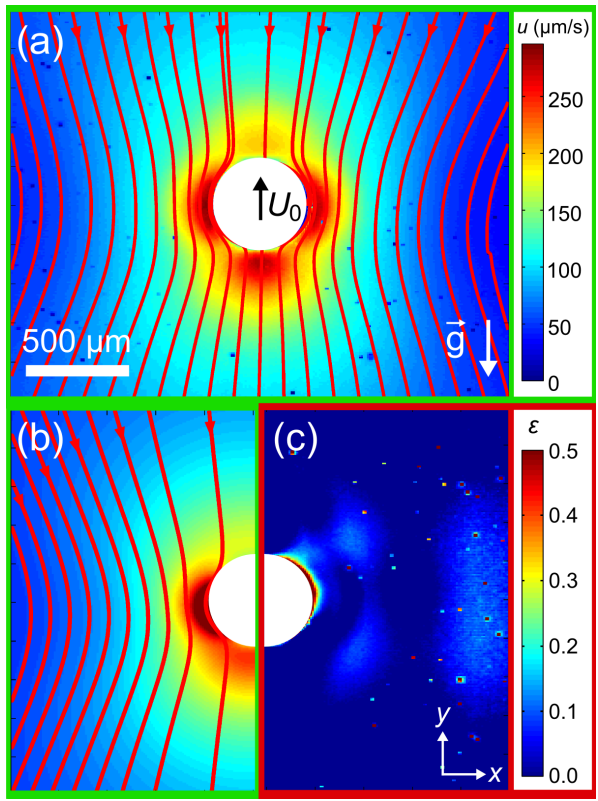}
\end{center}
\caption{\label{fig1} (color online). Flow field of a freely swimming {\it V.
carteri} in the laboratory frame. (a,b) Magnitude and streamlines of
$\mathbf{u}$ and its fitted approximation $\mathbf{u}_{fit}$ respectively. (c)
Relative error of the fit: $\epsilon = ||\mathbf{u}-\mathbf{u}_{fit}|| /
||\mathbf{u}||$. (a) and (b) have the same colorbar, different from (c).
$\vec{g}$ indicates gravity.}
\end{figure}

{\it V. carteri} f. {\it nagariensis} (strain EVE) was grown axenically in SVM
\cite{kirk83} with sterile air bubbling, whereas {\it C. reinhardtii} (strain UTEX
89) was grown axenically in TAP medium \cite{harris09} on an orbital shaker,
both in a diurnal growth chamber with $16\,$h in artificial cool daylight
($\sim4000\,$lux) at 28$^{\circ}\,$C, and $8\,$h in the dark at $26^{\circ}\,$C.
The large difference in organism size between {\it Volvox} and {\it
Chlamydomonas} required two distinct methods to measure the flows they create
\cite{SupplMat}. A CCD camera (Pike, Allied Vision Technologies) mounted on a
continuously-focusable microscope (Infinivar, Infinity Optics) and connected to
a vertical motorized XY stage (Thorlabs) followed individual {\it Volvox}
colonies as they swam upwards \cite{Volvox_waltzing} in a straight line along
the central axis of a $5\times5\times50\,$mm sample chamber filled with SVM at
$21\pm 1^{\circ}\,$C. The stage was controlled by a custom LabView routine. The
fluid was seeded at volume fraction $10^{-5}$ with $1\,\mu$m nile-red
polystyrene microspheres (Invitrogen) illuminated by a vertical $\sim 500\,\mu$m
thick laser sheet ($\lambda=532\,$nm). {\it Volvox} is phototactic
\cite{DrePNAS} at this wavelength, and at the intensities used here it swims
smoothly along the laser sheet. We recorded the flow field of $19$ different
colonies at $30\,$fps for $\sim 2-3\,$min each. The measured flow field
$\mathbf{v}$ was obtained by particle image velocimetry (Dantec Dynamics).
Background flows in the chamber were $<10\,\mu$m/s.

We observed a dilute suspension ($\sim 3\times 10^6$ cells/cm$^3$) of {\it
Chlamydomonas} in TAP on a Nikon inverted microscope at $40 \times$ (NA $0.6$)
by exciting their chlorophyll autofluorescence with a laser ($635\,$nm, $\sim
60\,$mW), which also excited $1.6\,\mu$m fluorescent polystyrene microspheres
(FS04F, Bangs Labs) used as tracers. Cylindrical polydimethylsiloxane sample
chambers ($5\,$mm radius, $0.4\,$mm height) were prepared, pacified, and filled
following \cite{Chlamytracers}. Experiments were performed at $21\pm
1^{\circ}\,$C, with the laser providing the only light source. We focused on a
plane $150\,\mu$m inside the chamber to minimize surface effects, and recorded
movies at $250\,$fps (Fastcam SA3, Photron). Movies were analysed with standard
algorithms to track cells and tracers. For each cell swimming along the focal
plane for more than $1\,$s ($\sim 10$ body lengths), we collected the
instantaneous velocity of all tracers at $r < 14 \,R$, normalized by the
swimmer's speed. The resulting $3.3 \times 10^6$ velocity vectors were binned
into a $2.5$ $\mu$m square grid (shown in Fig. \ref{fig4} below), and the mean
of the well-resolved Gaussian in each bin was used for the flow field.

In both experiments $\mathbf{U}_0$ indicates the swimmer velocity, while ${\bf
u}({\bf r})$ and ${\bf v}({\bf r})={\bf u}({\bf r})-\mathbf{U}_0$ are the
velocity field in the laboratory and comoving frames respectively.

\begin{figure}[t]
\begin{center}
\includegraphics*[clip=true,width=0.95\columnwidth]{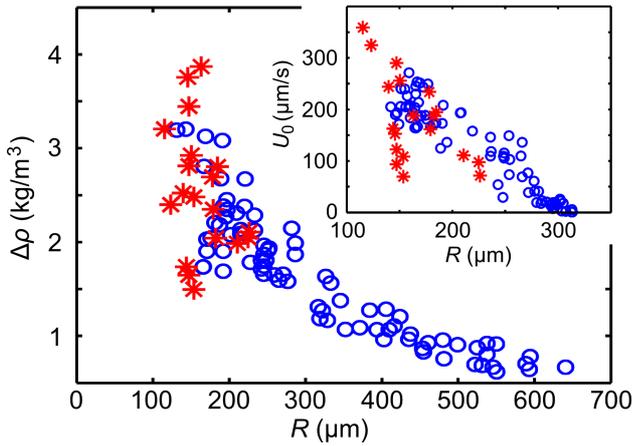}
\end{center}
\caption{\label{fig2} (color online). The dependence of the excess density
$\Delta\rho$ and swimming speed $U_0$ (inset) of {\it V. carteri} colonies (red
stars) on their radius $R$ are compatible with previous measurements (blue
circles) \cite{Volvox_waltzing}, and also \cite{SolariAmNat}.} 
\end{figure}

\begin{figure}[t]
\begin{center}
\includegraphics*[clip=true,width=0.90\columnwidth]{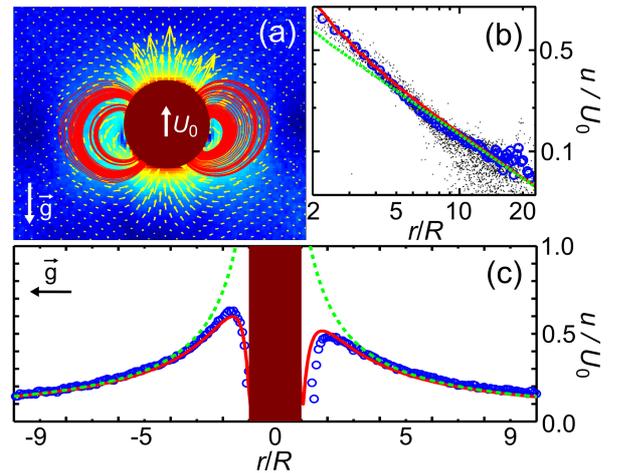}
\end{center}
\caption{\label{fig3} (color online). Near field around {\it V. carteri}. (a)
Magnitude, vector fields and streamlines of $\mathbf{u}$ after subtracting the
fitted Stokeslet. Colorbar as in Fig. \ref{fig1}(a). (b) $||\mathbf{u}||$ along
a horizontal section through the center of the organism. The average Stokeslet
(green dashed line) follows the decay of the experimental flow (blue circles)
averaged over 19 different colonies (black dots). Deviations from a pure
monopole appear from $\lesssim 5$R, and can be captured adding a source doublet
and a stresslet (red solid line). (c) Vertical section of the flow field
$\mathbf{u}$ from the experiment in (a) through the center of the colony. The
stresslet component is responsible for the forward-backward asymmetry. Symbols
as in (b).}
\end{figure}

A typical experimental flow field around {\it Volvox} is shown in Fig.
\ref{fig1}(a). We fit these fields to a superposition of a uniform background
velocity ($\mathbf{U}_0$), a Stokeslet (St), a stresslet (str) and a source
doublet (sd):
\begin{eqnarray}\label{eqn:vlvxflow}
\mathbf{v}_{fit}({\bf r})=
&-& U_0\,\hat{\bf y}-\frac{A_{St}}{r}\left(\mathbf{I}+\hat{\bf r}
\hat{\bf r}\right)\cdot\hat{\bf y} \\ 
&-&\frac{A_{str}}{r^2}\left(1-3(y/r)^2\right)\hat{\bf r} - 
\frac{A_{sd}}{r^3}\left(\frac{\mathbf{I}}{3}-\hat{\bf r}\hat{\bf r}\right)\cdot\hat{\bf y} \nonumber
\end{eqnarray}
where $\mathbf{I}$ is the unit tensor, $\hat{\bf y}$ is the upward vertical unit
vector, $\hat{\bf r}=\mathbf{r}/r$, and $\mathbf{\bf r}$ is measured from the
center of the organism $(x_c,y_c)$. The orientation of all multipoles is fixed
to be along the vertical, and we are left with six parameters:
$(U_0,A_{St},A_{str},A_{sd},x_c,y_c)$. The fits, obtained by minimizing the
integrated squared difference between the model and the experimental flows,
describe remarkably well the experimental flow, almost down to the surface of
the organisms [see Fig. \ref{fig1}(b,c)]. Typical values for the parameters are
$U_0\sim 10^2\,\mu$m/s, $A_{St}\sim 10^4\,\mu\textrm{m}^2$/s, $A_{str}\sim
10^6\,\mu\textrm{m}^3$/s (indicating a pusher-type stresslet),
$A_{sd}\sim10^9\,\mu\textrm{m}^4$/s, with the actual magnitude depending on the
colony radius $R$. From the Stokeslet component, we can calculate the average
colony density as $\Delta\rho=6\eta A_{St}/gR^3$, where
$\eta=10^{-3}\,\textrm{Pa}\,\textrm{s}$ and $g$ is the gravitational
acceleration.  The dependence of both $\Delta\rho$ and $U_0$ on $R$ (Fig.
\ref{fig2}) compares well with previously published data \cite{Volvox_waltzing}
obtained by different means, thereby validating the measurements and analysis
procedures. Removing the Stokeslet contribution from the experimental flow field
[Fig. \ref{fig3}(a)] reveals that the near field is dominated by the source
doublet component, with the stresslet responsible only for a slight
forward-backward asymmetry [Fig. \ref{fig3}(b,c)]. The orientation of the source
doublet is opposite to that around a translating solid sphere, and is compatible
with a model that assigns a constant force density to the colony surface
\cite{Short06}, as well as a particular case of the ``squirmer'' model
\cite{blake71b}. Average values of the parameters in Eq. \ref{eqn:vlvxflow} show
that the crossover distance between the source doublet and stresslet
($\sim3\,R$) is beyond that at which the Stokeslet becomes the leading component
of the flow ($\sim1.5\,R$). This peculiar ordering of multipoles results from
the high degree of anterior-posterior symmetry of {\it Volvox} \cite{Short06},
and highlights the influence that a swimmer's body plan can have on its flow
field.

\begin{figure*}[t]
\begin{center}
\includegraphics*[clip=true,width=1.8\columnwidth]{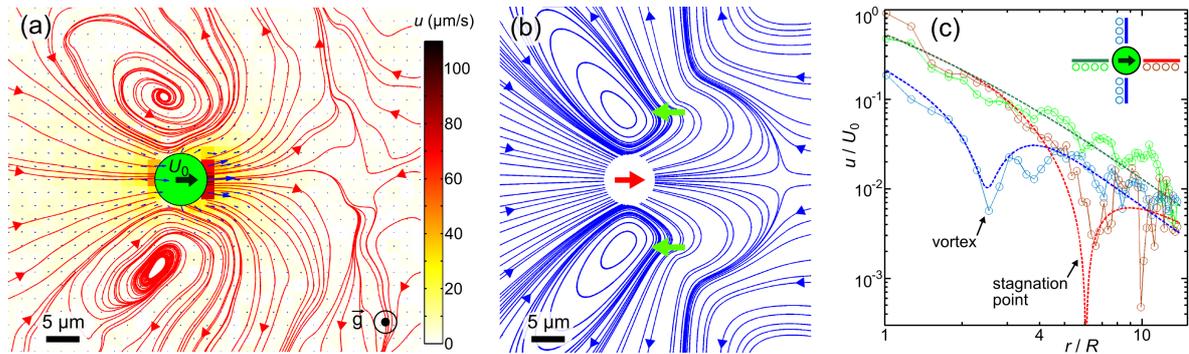}
\end{center}
\caption{\label{fig4} (color online). Time- and azimuthally-averaged flow field
of {\it C. reinhardtii}. (a) Streamlines (red) computed from velocity vectors
(blue). 
The spiraling near elliptic points is an artifact of the direct integration
a noisy experimental velocity field.
A color scheme indicates flow speed magnitudes. (b) Streamlines of the
azimuthally averaged flow of the three-Stokeslet model: flagellar thrust is
distributed among two Stokeslets placed (not fitted) at the approximate
flagellar position (lateral green arrows), whose sum balances drag on the cell
body (central red arrow). (c) Decay of $||\mathbf{u}(\mathbf{r})||$ for the
three directions indicated by separate colors in the inset, compared to results
from the three-Stokeslet model (dashed).} 
\end{figure*}

The two $\sim 12\,\mu$m long flagella of {\it Chlamydomonas} beat mostly in a
synchronous breast stroke at $\sim 50\,$Hz \cite{Chlamy_Science}, pulling the
cell body through the fluid at speeds $U_0 \sim 100\,\mu$m/s. Despite its $\sim
5$\% density excess over water \cite{SolariAmNat}, the gravitational Stokeslet
of {\it Chlamydomonas} only becomes dominant at distances $\Lambda \gtrsim 35
\,R$, as its ratio of $U_0$ to the sedimentation speed ($F_g / 6 \pi \eta R$)
is much larger than for {\it Volvox}. Therefore this swimmer has often been
modelled as a puller stresslet \cite{PedleyKessler}. A slight
three-dimensionality of the beating causes the cell to spin about its swimming
direction at $\sim 2\,$Hz, so ensemble-averaged measurements of the kind
presented here average out azimuthal asymmetries in the flow field.  Figure
\ref{fig4}(a) shows that for $r \gtrsim 7R$ the measured flow topology begins to
resemble a puller stresslet, yet flow speeds at such distances are already
$\lesssim 1$ $\mu$m/s. Closer to the organism, the field becomes more complex.
It includes side vortices and a flow in front of the cell body that is {\it
along} the direction of motion, towards a stagnation point. The velocity field
can be accurately captured, even in the near-field, by modeling the flow created
by the pulled cell body as a Stokeslet, distributing the thrust among two
Stokeslets located at the approximate positions of the two flagella, and
averaging the flow over one rotation about $\mathbf{U}_0$. The streamlines of
this simple extension to the force dipole model [see Fig. \ref{fig4}(b)], as well as
the decay of $||{\bf u}({\bf r})||$ with distance [see Fig. \ref{fig4}(c)], are very
similar to those measured. Including no-slip boundary conditions on the cell
body \cite{oseen} has little effect on the velocity field as the cell-drag
Stokeslet nearly produces the appropriate velocity field on the cell surface. 

These flow field measurements around freely-swimming microorganisms provide the
basis for a deeper understanding of a number of issues in biological fluid
dynamics, including the interactions of microorganisms with surfaces, with each
other, and the rheology of suspensions. For example, it was recently discovered
that {\it Volvox} colonies can form hydrodynamic bound states whose properties
are quantitatively described by a model of interacting Stokeslets near a no-slip
wall \cite{Volvox_waltzing}. The near complete dominance of the flow field
around {\it Volvox} by the Stokeslet term found here provides \textit{ex post
facto} justification for the neglect of higher moments.  Perhaps more
importantly this result shows that in terms of interparticle hydrodynamic
couplings a suspension of {\it Volvox} is like a sedimenting suspension
\cite{sedimentation}, except that the velocity of each colony is the sum of a
self-propelled contribution and mutual advection in the flow field of other
spheres. Elsewhere we illustrate this correspondence in detail \cite{Kantsler}.

The correspondence between the measured time- and azimuthally-averaged flow
field of {\it Chlamydomonas} and the three-Stokeslet model illustrates how well
such a simplification captures the complex flow topology, lending support to
this approximation in modelling ciliary interactions \cite{point_cilia}. Our
results indicate that the simple puller-type description for {\it Chlamydomonas}
is only valid at distances $\gtrsim 7\,R$, where the flow field is already
$\lesssim 1\%$ of $U_0$. We then expect interactions with other swimmers,
boundaries or tracers, to be influenced mostly by the flow structure at shorter
separations, where the full time dependence of the flow may be important \cite{GG}. 
We are currently investigating whether similar conclusions hold for the
flow field around bacteria, the prototypical ``pusher''
microorganisms.

We thank K.C. Leptos for suggesting the use of autofluorescence to track {\it
Chlamydomonas} cells, S.B. Dalziel, V. Kantsler and T.J. Pedley for discussions, D. Page-Croft and
N. Price for technical assistance, and acknowledge support from the EPSRC, the
BBSRC, the Marie-Curie Program (M.P.), and the Schlumberger Chair Fund.

\end{document}